\documentstyle[psfig,conf_iap,]{article}
\begin{document}
\heading{%
%
Modelling Quintessential Inflation with Branes
%
}
\par\medskip\noindent
\author{%
Konstantinos Dimopoulos$^{1,2}$
}
\address{%
Physics Department, Lancaster University, Lancaster LA1 4YB, UK
}
\address{%
Department of Physics, University of Oxford,
Keble Road, Oxford OX1 3RH, UK
}

\begin{abstract}
I discuss why quintessential inflation model-building is more natural in the
context of brane cosmology and study a particular model as an example.
\end{abstract}
\newcommand{\Gsim}{\mbox{\raisebox{-.9ex}{~$\stackrel{\mbox{$>$}}{\sim}$~}}}
\section{Introduction}
To minimize its fine-tuning problems, such as initial conditions, there have
been attempts to unify quintessence with the inflaton field in a single
scalar \cite{mine,multi1,multi2,multi3}. This way, introducing yet again
another unobserved scalar field is avoided. Also, a common theoretical
framework can be used to describe both inflation and quintessence. For
quintessential inflation one needs a sterile inflaton, with only minimal
gravitational coupling to the standard model, because the field should survive
until today. The Universe reheats through gravitational particle production
\cite{ford}. The minimum of the potential (assumed zero) is typically placed
at infinity, because it should not have yet been reached. This results in the
so-called quintessential tail of the potential. Candidates for the
quintessential--inflation scalar field are moduli fields or the radion field.
\nopagebreak[4]
\section{Dynamics}
The Universe is modeled as a collection of perfect fluids; the background fluid
with density $\rho_B$ (comprised by matter and radiation) and the scalar field
$\phi$ with density $\rho_\phi=\rho_{\rm kin}+V$ and pressure
$p_\phi=\rho_{\rm kin}-V$, where $\rho_{\rm kin}\equiv\frac{1}{2}\dot{\phi}^2$
is the kinetic density of $\phi$ and the dot denotes derivative w.r.t. the
cosmic time $t$. For every component one defines the barotropic parameter as
\mbox{$w_i\equiv p_i/\rho_i$}. The Universe expansion accelerates when
$\rho_B<\rho_\phi$ and $w_\phi<-\frac{1}{3}$. Energy-momentum conservation
demands \mbox{$d(a^3\rho) = -p\,d(a^3)$}, which, for decoupled fluids, gives
\mbox{$\rho_i\propto a^{-3(1+w_i)}$}, where $a$ is the scale factor. To study
the dynamics of the Universe, one also needs the Friedman equation and the
$\phi$ field equation:
\mbox{$\dot{\rho}_{\rm kin}+6H\rho_{\rm kin}+\dot{V}=0$}, where
\mbox{$H\equiv\dot{a}/a$} is the Hubble parameter.
In spatially-flat, FRW cosmology the Friedman equation is
\mbox{$H^2 = \rho/3m_P^2$}, with $m_P$ being the reduced Planck mass. This
results in the evolution equations:
\begin{eqnarray}
H=\frac{2\,t^{-1}}{3(1+w)} &
\qquad a\propto t^{\frac{2}{3(1+w)}}\qquad  &
\rho=\frac{4}{3(1+w)^2}\Big(\frac{m_P}{t}\Big)^2
\end{eqnarray}
where $w$ corresponds to the dominant fluid component.
In brane cosmology the Friedman equation is modified for density above
the string tension $\lambda$ \cite{dnmics}. Assuming a 5th dimension, the
Friedman equation becomes \mbox{$H\simeq\rho/\sqrt{6\lambda}\,m_P$}, where
\mbox{$\lambda=\frac{3}{4\pi}(M_5^6/m_P^2)$}, with $M_5$ being the
fundamental, 5-dim Planck mass. Then the evolution equations are:
\begin{eqnarray}
H=\frac{t^{-1}}{3(1+w)} &
\qquad a\propto t^{\frac{1}{3(1+w)}}\qquad  &
\rho=\frac{\sqrt{6\lambda}}{3(1+w)}\Big(\frac{m_P}{t}\Big)
\end{eqnarray}
The modified dynamics affect the inflationary era due to excessive friction on
the roll of the scalar field, which allows inflation with steep potential.
Then, the slow-roll parameters become:
\mbox{$\epsilon\simeq 2\lambda m_P^2(V')^2/V^3$} and
\mbox{$\eta\simeq 2\lambda m_P^2V''/V^2$},
where the prime denotes derivative w.r.t. $\phi$. The modified slow-roll
changes the amplitude and spectral index of the generated density
perturbations \cite{dnmics}:
\begin{eqnarray}
\frac{\delta\rho}{\rho}\simeq
\frac{1}{2\sqrt{6}\,\pi}\frac{V^3}{\lambda^{3/2}|V'|m_P^3}
& \mbox{and} &
 n_s-1\simeq -4m_P^2\frac{\lambda}{V}
\left[\;3\left(\frac{V'}{V}\right)^2-\frac{V''}{V}\;\right]\quad
\end{eqnarray}
Gravitational reheating creates a thermal bath of temperature
\mbox{$T_{\rm reh}=\frac{\alpha}{2\pi}\,H_{\rm end}$} \cite{ford}, where
$\alpha$ is the reheating efficiency and the subscript 'end' denotes
the end of inflation. Due to the inefficiency of reheating, after the end
of inflation the Universe becomes dominated by $\rho_{\rm kin}$, i.e.
\mbox{$\rho\simeq\rho_{\rm kin}\propto a^{-6}$}. As long as
\mbox{$\rho_{\rm kin}>2\lambda$} we have \mbox{$a\propto t^{1/6}$} and
\mbox{$\phi(t)=\phi_{\rm end}+\frac{2}{\sqrt{3}}\sqrt{\lambda/V_{\rm end}}
\left(\sqrt{t/t_{\rm end}}-1\right)m_P$}. However, when
\mbox{$\rho_{\rm kin}<2\lambda$} the brane regime ends and one has
\mbox{$a\propto t^{1/3}$} and \mbox{$\phi(t)=\phi_{\rm end}+\frac{1}{\sqrt{6}}
\left[2+\ln\left(t/t_\lambda\right)\right]m_P$}, where
\mbox{$t_\lambda=\frac{1}{2\sqrt{6}}\,m_P/\sqrt{\lambda}$} is the crossover
time for which \mbox{$\phi_\lambda=\phi_{\rm end}+\sqrt{2/3}\;m_P$}. The
thermal bath eventually dominates the density of the Universe and the Hot Big
Bang begins at the temperature:
\begin{equation}
T_*=\frac{\alpha^3}{96\pi^2}\sqrt{\frac{g_*}{15}}
\frac{V_{\rm end}^{5/2}}{\lambda^{3/2}m_P^3}
\end{equation}
where $g_*\sim 10^2$ is the number of relativistic degrees of freedom. In
order not to affect Nucleosynthesis (BBN) we require \mbox{$T_*> 1$ MeV}.
After the onset of the Hot Big Bang $\rho_{\rm kin}$ rapidly
decreases and the field freezes at the value:
\begin{equation}
\phi_F=\phi_{\rm end}+\frac{m_P}{\sqrt{6}}\left[
14+\frac{3}{2}\ln\left(\frac{30\pi^2}{g_*\alpha^4}\right)-
4\ln\left(\frac{m_P^4}{\lambda}\right)
+5\ln\left(\frac{m_P^4}{V_{\rm end}}\right)
\right]\end{equation}
\section{Why Branes}
Standard--cosmology quintessential inflation needs a potential with two flat
regions: the inflationary plateau and the quintessential tail \cite{mine}. BBN
and coincidence demand that the density scale of the flat regions differs by
$\sim 10^{100}$! To prepare for this abysmal dive, $V$ is strongly curved
near the end of inflation. As a result, the slow-roll parameter $\eta$ is
large, leading to spectral index $n_s$  far from unity. This makes it
difficult to construct single-branch models\footnote{Multi-branch models such
as \cite{multi1,multi2,multi3} are disfavoured in our minimalistic approach.}
for quintessential inflation, although solutions do exist \cite{mine}.
In contrast, this $\eta$-problem does not appear when considering brane
cosmology because the modified Friedman equation allows for steep inflation
\cite{steep,lidsey,brane1,dnmics,brane2}. Hence, one may obtain
successful quintessential inflation with a simple potential.
\section{The exponential tail}
String theory disfavours eternal acceleration \cite{string1,string2}. This
immediately rules out quintessential tails milder than exponential and also
frozen quintessence (in which \mbox{$\phi=\phi_F$} at present). In any case,
coincidence is hard to achieve with mild tails (e.g. inverse power-law type
\cite{lidsey}) and frozen quintessence is virtually indistinguishable from the
cosmological constant alternative. On the other hand steeper-than-exponential
tails have disastrous attractors \cite{mine}. Thus, the most reasonable
approach is the exponential tail, \mbox{$V\simeq V_0e^{-b\phi/m_P}$}, where
$b$ is a positive constant. Then, the $\phi$-field equation gives an exact
attractor solution:
\begin{eqnarray}
\phi_{\rm attr}=\frac{2m_P}{b}
\ln\!\left[\sqrt{\frac{V_0}{2}\left(\frac{1+w}{1-w}\right)}\,
\frac{b\,t}{m_P}\right] & \mbox{and} &
V_{\rm attr}=\frac{2}{b^2}\left(\frac{1-w}{1+w}\right)
\left(\frac{m_P}{t}\right)^2\qquad
\end{eqnarray}
If $\rho_\phi<\rho_B$ then \mbox{$w=w_B$} and
\mbox{$\rho_\phi/\rho_B=3(1+w_B)/b^2=$ const.}, which means
\mbox{$w_\phi=w_B\geq 0$} and \mbox{$b^2>3(1+w_B)$}.
If, however, $\rho_\phi>\rho_B$ then \mbox{$w=w_\phi$} and
\mbox{$\rho=\rho_\phi\propto a^{-3(1+w_\phi)}$}.
This time \mbox{$b^2=3(1+w_\phi)$}. Therefore, {\em dark energy dominates
without eternal acceleration when} \ \mbox{$2<b^2< 3(1+w_B)$}.
Brief acceleration is possible at the time when $\phi$ unfreezes to follow the
attractor, due to superfreezing \cite{mine}. In fact, superfreezing enlarges
the range to \mbox{$2<b^2<24$} \cite{cline}.
\section{A toy-model example}
\mbox{Consider the potential (also studied in \cite{rachel}):}
\begin{equation}
V(\phi)=\frac{M^4}{\cosh(b\phi/m_P)-1}\quad\Rightarrow\quad
\left\{\begin{array}{ll}
V\simeq 2M^4e^{-b\phi/m_P} & \phi\gg m_P/b\\
V\simeq 2M^4(m_P/b\phi)^2 & \phi\ll m_P/b
\end{array}\right.
\end{equation}
The slow-roll parameters are:
\mbox{$\eta=2A[\cosh(b\phi/m_P)+2]=\epsilon+2A$}. Thus, inflation
ends when \mbox{$\eta(\phi_{\rm end})\equiv 1$}, where
\mbox{$A\equiv b^2(\lambda/M^4)$}. The spectral index is
\begin{equation}
n_s-1=-4A\;\frac{3+\left(\frac{1-6A}{1-2A}\right)\exp(-4N_{\rm dec}A)}{1-
\left(\frac{1-6A}{1-2A}\right)\exp(-4N_{\rm dec}A)}
\end{equation}
where \mbox{$N_{\rm dec}\simeq 69$} is the number of e-folds of remaining
inflation corresponding to the scale of the horizon at decoupling. From the
observations \mbox{$|n_s-1|\leq 0.1$}, which demands the constraint
\mbox{$A<\frac{1}{148}$}. Using \mbox{$A\ll 1$} the above becomes
\mbox{$n_s-1\simeq-\frac{4}{N_{\rm dec}+1}$}, which gives
\mbox{$n_s\simeq 0.94$}. The density contrast with \mbox{$A\ll 1$} is:
\begin{equation}
\frac{\delta\rho}{\rho}\simeq\frac{2b^2}{\sqrt{6}\pi}
\left(\frac{M}{m_P}\right)^2\sqrt{A}\,(N_{\rm dec}+1)^2
\end{equation}
Similarly, for the inflationary scale we find \mbox{$V_{\rm end}\simeq 2AM^4$}
or, equivalently,
\begin{equation}
V_{\rm end}\simeq
\frac{3\pi^2}{b^4}\left(\frac{\delta\rho}{\rho}\right)^2
\frac{m_P^4}{(N_{\rm dec}+1)^4}
\label{Vend}
\end{equation}
The field freezes at
\mbox{$\phi_F=\phi_{\rm end}+\frac{1}{\sqrt{6}}(61.7-4\ln b)m_P$}. Then
coincidence demands: \mbox{$b\simeq 14.5$}, which is too large for brief
acceleration. Further, Eq.(\ref{Vend}) gives
\mbox{$V_{\rm end}^{1/4}\simeq 2\!\times\!10^{13}$GeV}. Using
\mbox{$V_{\rm end}\simeq 2b^2\lambda$} one finds
\mbox{$\lambda^{1/4}\simeq 6\!\times\! 10^{12}$GeV} and so
\mbox{$M_5\simeq 5\!\times\!10^{14}$GeV}. The bound on $A$ suggests:
\mbox{$M>7\!\times\! 10^{13}$GeV}, which enables the identification
\mbox{$M=M_5$}. The Hot Big Bang begins at temperature:
\begin{equation}
T_*=\frac{\alpha^3}{16b(N_{\rm dec}+1)^4}\sqrt{\frac{2g_*}{15}}
\left(\frac{\delta\rho}{\rho}\right)^2m_P
\end{equation}
which satisfies the BBN constraint if \mbox{$\alpha\Gsim 0.1$}.
With \mbox{$\alpha\sim 0.1$},
the modified Friedman equation gives
\mbox{$T_{\rm reh}\sim 10^7$GeV},
which satisfies the gravitino bound.
\section{Conclusions}
Quintessential-inflation model-building is easier in brane-cosmology because
the $\eta$-problem is overcome by considering steep inflation \cite{steep}.
The dynamics of the Universe from inflation through to the present have been
analysed and employed in the toy-model presented. This model incorporates two
natural mass-scales ($m_P$ and $M_5$) and leads to a spectral index within
observational bounds. However, it fails to provide late-time
accelerated expansion.

\acknowledgements{Work supported by the EU network: {\sc hprn-ct-00-00152}.}

\vfill
\end{document}